# Janus-graphene: a two-dimensional half-auxetic carbon allotropes with non-chemical Janus configuration

Linfeng Yu[1], Jianhua Xu[1], Chen Shen[2], Hongbin Zhang[2], Xiong Zheng[1], Huiming Wang[3], Zhenzhen Qin[4] and Guangzhao Qin[1,5,*]

[1]*National Key Laboratory of Advanced Design and Manufacturing Technology for Vehicle, College of Mechanical and Vehicle Engineering, Hunan University, Changsha 410082, P. R. China*

[2]*Institute of Materials Science, Technical University of Darmstadt, Darmstadt 64287, Germany.*

[3]*Hunan Key Laboratory for Micro-Nano Energy Materials & Device and School of Physics and Optoelectronics, Xiangtan University, Xiangtan 411105, Hunan, China*

[4]*School of Physics and Microelectronics, Zhengzhou University, Zhengzhou 450001, China*

[5]*Greater Bay Area Institute for Innovation, Hunan University, Guangzhou 511300, Guangdong Province, P. R. China*

---

[*] Author to whom all correspondence should be addressed. E-Mail: gzqin@hnu.edu.cn

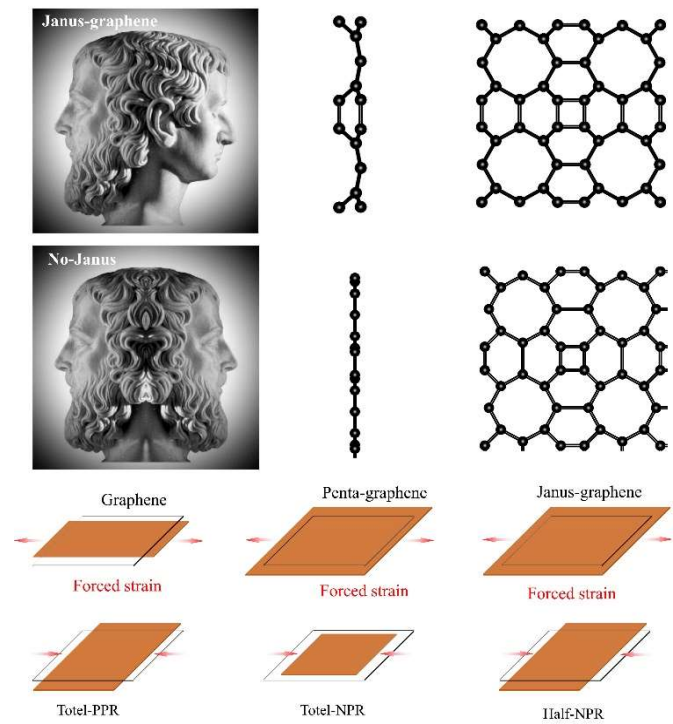

Janus is the god of heaven in Roman mythology. He has a face on the front and back of his head, also known as the two-faced god, who guards the door and the beginning and end of all things. Here we report a novel carbon allotrope with the Janus configuration called as Janus-graphene, which exhibits bidirectional and bifunctionality along the *out-of-plane* direction. An unconventional half-auxetic behaviour is found in the Janus-graphene, *i.e.,* it always expand whether stretched or compressed.

**Abstract**

The asymmetric properties of Janus two-dimensional materials commonly depend on chemical effects, such as different atoms, elements, material types, *etc*. Herein, based on carbon gene recombination strategy, we identify an intrinsic non-chemical Janus configuration in a novel purely *sp*$^2$ hybridized carbon monolayer, named as Janus-graphene. With the carbon gene of tetragonal, hexagonal, and octagonal rings, the spontaneous unilateral growth of carbon atoms drives the non-chemical Janus configuration in Janus-graphene, which is totally different from the chemical effect in common Janus materials such as MoSSe. A structure-independent half-auxetic behavior is mapped in Janus-graphene that the structure maintains expansion whether stretched or compressed, which lies in the key role of $p_z$ orbital. The unprecedented half-auxeticity in Janus-graphene extends intrinsic auxeticity into pure *sp*$^2$ hybrid carbon configurations. With the unique half-auxeticity emerged in the non-chemical Janus configuration, Janus-graphene enriches the functional carbon family as a promising candidate for micro/nanoelectronic device applications.

# 1. Introduction

Carbon element has extremely strong bonding ability and abundant bonding forms in chemical reactions, such as linear alkyne (*sp*), planar olefin (*sp*$^2$), and tetrahedral alkane bonds (*sp*$^3$). The superior performance of graphene, such as superconducting[1–3], and quantum Hall effect[4,5], motivate researchers to extend their interest from three-dimensional to two-dimensional (2D) carbon crystals. Many carbon monolayers with promising properties have been theoretically predicted or experimentally synthesized, including penta-graphene[6], T-graphene[7], monolayer amorphous carbon[8], biphenylene[9], graphene+[10], fullerene monolayer[11], and others[12–19]. The discovery of these promising carbon configurations has encouraged the designing and screening of multifunctional carbon monolayers with specific technological interests.

With the development of computer alchemy, many crystal structure prediction methods can be served as valid candidates for designing and screening novel carbon configurations, such as graph theory[20,21], particle swarm methods[22], simulated annealing[23], genetic algorithm[24], and other methods[25,26]. Those methods facilitate the discovery and exploration of novel potential functional carbon configurations. However, the successful design relies on a clear understanding of the bonding morphology and geometric features, which allow the identification of unique topological effects as effective guides for designing promising carbon monolayers beyond graphene. A clear insight into the C-C bonding morphologies and their geometric features is urgently needed and will accelerate the efficient design and screening of carbon materials with multifunction.

As an emerging class of nanomaterials, Janus nanostructures are functional structures with different properties composed of two materials (or different elements) on the same object, driven by chemical effects. "Janus bead" was first proposed by French scientist Casagrande in 1989 to describe the unique structure of semi-hydrophobic and semi-hydrophilic on the surface of spherical particles[27]. Then, in 1991, the name "Janus" of the two-faced god in Greek mythology was first proposed by French physicist Gennes to describe particles with dual properties in his Nobel Prize speech[28]. Subsequently, Janus materials have been widely used in many fields, such as biology, energy, chemistry, and physics, due to their unique asymmetry[29–37].

2D Janus atomic crystals are novel monolayer materials with structural symmetry breaking caused by chemical effects. Tremendous effort has been devoted to designing and synthesizing the 2D Janus material with chemisorption[38,39], atomic substitution[33], and stacked heterojunctions[40], facilitating the generation of many 2D nanomaterials with unique physical properties, such as MoSSe[33]. Note that the Janus properties of current Janus materials depend on chemical effects, *i.e.*, different adatoms, functional groups, or elemental effects. Currently, the preparation of Janus materials relies on chemical interactions, including atomic substitution[33,41,42], chemisorption[38,39,43,44], and heterojunction[40,45,46], which divides Janus materials into three types as shown in Fig. 1: (1) The element-type Janus structure, where different elements are arranged along the out-of-plane direction, such as 2H-MoSSe[33]; (2) The adatom-type Janus structure, which are formed by the adsorption of different atoms on both sides along the out-of-plane sides, such as graphene-like Janus materials[47]; (3) The heterojunction-type Janus structure, which are van der Waals layered compounds formed by different types of monolayers along out-of-plane directions, such as graphene/MgX (X=S, Se) and graphene/BN[48–51]. However, intrinsic carbon materials, where only carbon atoms exist, can hardly achieve Janus properties without extrinsic chemical effects. Such a problem inspires us to design 2D Janus carbon materials with unique symmetry breaking using advanced crystal design methods.

In this work, we propose a carbon gene reconstruction strategy to assist the design of 2D carbon materials by establishing a carbon gene bank. With this strategy, we demonstrate a successful design for a novel 2D carbon allotrope with $sp^2$ hybridization, *i.e.*, Janus-graphene. The Janus-graphene exhibits an unprecedented non-chemical Janus configuration, which is totally different from the previously reported Janus structures driven by chemical effects. An emerging half-auxetic property is found in the pure $sp^2$-hybrid Janus-graphene, which breaks the notion that pure $sp^2$-hybridized carbon monolayers cannot exhibit intrinsic NPR behavior. This bottom-up design strategy provides a clear image of the bonding topography for understanding the carbon crystal configuration. The successfully designed Janus-graphene exhibits its unique Janus structural and unconventional half-auxeticity, which would be a strong candidate for the applications of carbon materials in micro/nano electronic devices.

## 2. Results and Discussion

### 2.1 Carbon Gene Recombination

Based on 2D carbon geometry features, a carbon gene recombination strategy can be proposed here to design 2D carbon with different configurations, which maps the geometric properties into quotient maps for recombination and screening of 2D carbon structures. As shown in Fig. 1(a), the initial idea of the strategy is to build a carbon gene bank according to the bonding characteristics of carbon materials. According to the number of atoms, 2D carbon genes can be divided into monoatomic chain, diatomic chain, triangular ring, tetragonal ring, pentagonal ring, hexagonal ring, *etc.*, as shown in Fig. 1 (b). In the *xy* plane, 2D carbon can be broken down into these carbon genes, and likewise, it can also be reconstituted to form novel carbon monolayers. As shown in Fig. 1(c), four hexagonal rings are reconstructed along the tetragonal rings, and then octagonal rings are introduced by periodic mirroring in the tetragonal lattice, forming a novel carbon monolayer with *sp*$^2$ hybridization.

The ground state configuration is the space with the lowest potential energy surface following the principle of energy minimization. Here, the initial structure with different degrees of deformation is randomly (or subjectively) perturbed to explore a sufficiently rich phase space, contributing to a relatively optimal solution, as shown in Fig. 1(c-d). In general, molecular dynamics can be more efficient than first-principles methods, but this requires high-precision potential functions, which can be facilitated by machine learning potentials[52]. Inspired by graphene, planar structures may be considered relatively more stable configurations. However, the buckling phase is found to have a lower potential energy surface than the planar phase in Janus-graphene, as shown in Fig. 1(d), implying stronger energy stability. Such evolution can also be used to reconstruct other 2D carbon configurations with specific carbon genes, as shown in Fig. 1(e)[53,54].

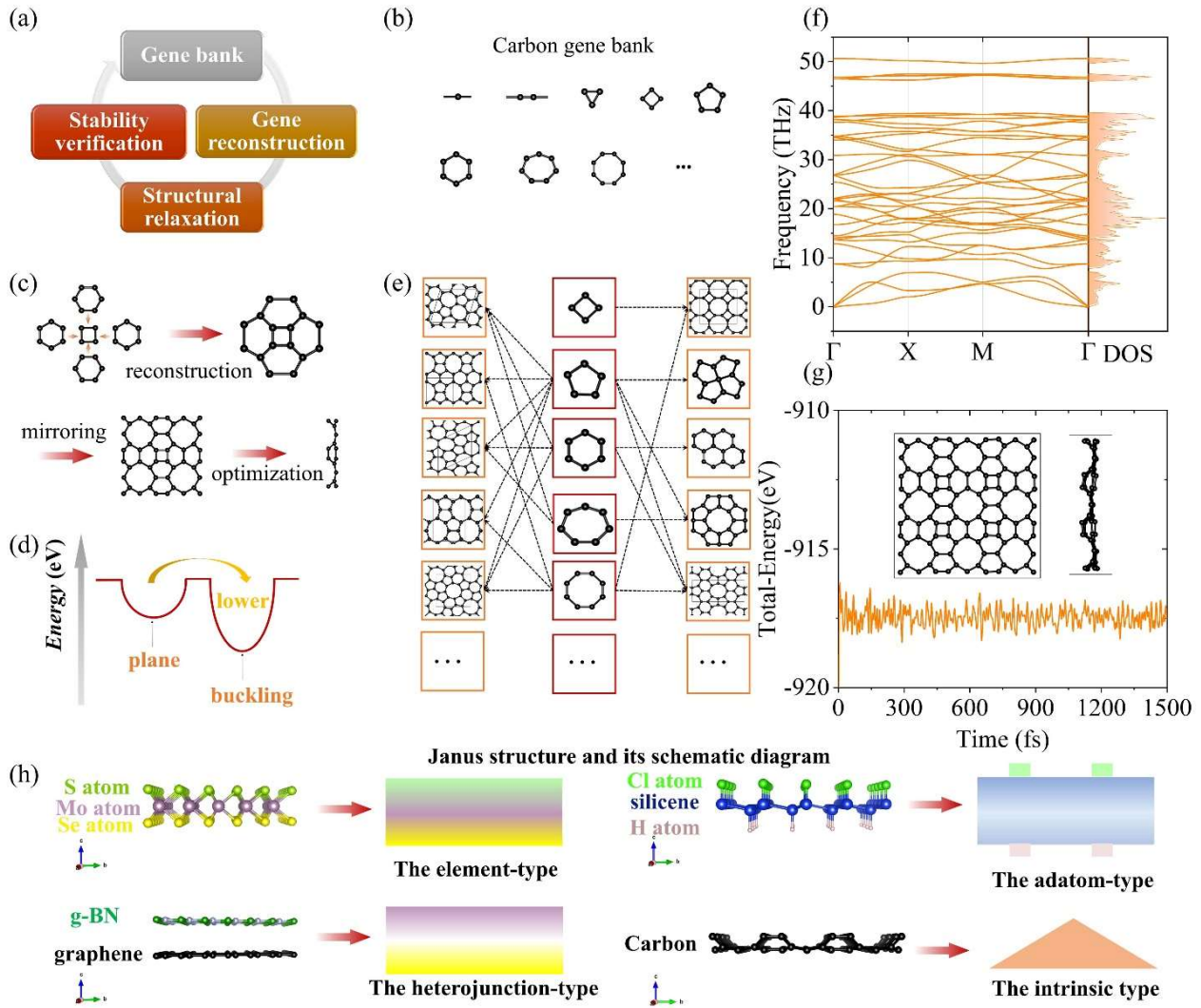

Figure 1. Janus-graphene and its evolution. (a) Carbon gene recombination workflow. (b) Two-dimensional carbon structure gene bank. (c) Geometric reconstruction of Janus-graphene. (d) Schematic diagram of the potential well distribution of the planar and buckled phases for Janus-graphene. (e) Two-dimensional carbon structure constructed by carbon gene recombination. (f) The phonon dispersion and phonon density of states (DOS) of Janus-graphene. (g) *Ab initio molecular dynamics simulations* (AIMD) at room temperature. (h) Janus structure type, including the element-type Janus structure and take 2H-MoSSe as an example, the adatom-type Janus structure and take silicene-like $Si_2ClH$ as an example, the heterojunction-type Janus structure and take graphene/g-BN heterojunction as an example, and intrinsic Janus structure in Janus-graphene.

## 2.2 Non-chemical Janus structure

The primitive cell of the optimized structure contains 12 atoms with the space group of P4mm (99). Phonon dispersion visualizes lattice vibrations in the Brillouin zone, reflecting the dynamic stability of the system as shown in Fig. 1(f), where the lowest-frequency phonon branches are all positive frequencies. Further, *ab initio molecular dynamics* simulations are performed to evaluate the thermodynamic stability at finite temperatures. As shown in Fig. 1(g), system energy can be conserved during thermal motion and the simulated structure at 300 K does not undergo cracking reactions, indicating thermal stability at room temperature.

Interestingly, this novel 2D carbon allotrope protrudes along one side of out of plane but exhibits planar features on the other side, which symbolizes the two-faced god Janus in ancient Roman mythology, hence the name "Janus-graphene". In carbon material, graphene can be formed with the above-mentioned Janus carbon materials through asymmetric chemistry[38,39,43,44], but usually leads to some irreversible changes in physical and chemical properties, such as the destruction of the Dirac cone[38]. Unlike chemically driven Janus materials, the Janus configuration of Janus-graphene is natural and intrinsic, as it purely consists of carbon atoms. Thus, it is considered as a new class of intrinsic Janus structure type. Unlike the Janus material summarized in Fig. 1(h), the intrinsic Janus structure relies on an uneven distribution of the structure in space, *i.e.*, the carbon atoms break the plane symmetry through an inhomogeneous arrangement to form the Janus phase, which leads to bifunctionality on both sides in the out-of-plane direction.

**2.3 Emerging half-auxeticity**

The unique Janus configuration brings extraordinary structural evolution performance, making its mechanical properties worthy of further study, especially the auxetic behavior. As shown in Fig. 2, diverse auxetic behaviors are found in Janus-graphene compared to typical graphene and penta-graphene. Graphene exhibits a total positive Poisson's ratio (Total-PPR) behavior in that it contracts (stretches) in orthogonal directions when stretched (compressed), as shown in Fig. 2(a), consistent with previous study[55]. Unlike graphene, a total negative Poisson's ratio (Total-NPR) behavior is found in penta-graphene as shown in Fig. 2(b), *i.e.*, it exhibits stretch (compression) behavior when stretched (compressed) in orthogonal directions[6]. Based on auxetic properties, penta-graphene can be an excellent platform to design a series of auxetic penta-materials through element replacement[56–60]. Such a phenomenon has also been intensively reported in hinge structures[61–63], TMCs[64], and other 2D configurations[65–69]. Unusually, Janus-graphene always expands regardless of whether it is stretched or compressed, *i.e.*, half-negative Poisson's ratio (Half-NPR) behavior, as shown in Fig. 2(c). The in-plane half-NPR behavior was first proposed by Ma *et al.*, but so far only been found in borides[66,70]. Unlike traditional auxetic behavior that the material expands (shrinks) when stretched (compressed), half-auxetic behavior reveals a mechanical phenomenon whereby materials always expand, whether stretched or compressed. Schematic representations of diverse Poisson's ratio behaviors are further plotted in Fig. 2(d-e) to understand their differences, *i.e.*, total-PPR for graphene, total-NPR for penta-graphene, and half-NPR for Janus-graphene. Half-NPR behavior of Janus-graphene is intrinsic and only occurs near the equilibrium position, implying that such a behavior can be successfully captured without applying additional means and fields. While this behavior extended in the out-of-plane direction in our previous study[10], no in-plane behavior was found in 2D carbon structures.

In penta−graphene[71], tetrahex-carbon[72], traditional auxetic behavior can be reproduced by elemental recombination, but it's hard to be reproduced in the derivatives of Janus-graphene. To further demonstrate the scarcity and magic of half-NPR behavior, we explored the Poisson's ratio behavior of monadic, binary, and ternary phase Janus materials via elemental recombination, as shown in Fig. 2(g–i). Note that only Janus-graphene expands when stretched, while all other materials contract, revealing its "uniqueness". The auxetic behavior is unique because low-dimensional quantum effects dominate the geometric evolution behavior at the microscopic level, which has been explained in detail in our previous studies[73–75]. Furthermore, the current auxetic behavior is only found in the $sp^3$ (or $sp^2$-$sp^3$) hybrid configuration in carbon materials. We searched the 108 carbon structures included in the 2D carbon database[76] and the 1114 $sp^2$-hybridized 2D carbon structures screened by He *et al.*[77], but found no auxetic behavior reported in the pure $sp^2$-hybridized carbon configuration. The reason lies in that the strong $sp^2$ hybridization favors the graphene-like planar structure, which hinders the generation of the reentry mechanism that induces NPR behavior. Here, due to the evolution of the unique Janus structure, the NPR behavior is extended to pure $sp^2$-hybrid carbon configurations, and this in-plane half-auxetic behavior is revealed for the first time in carbon materials.

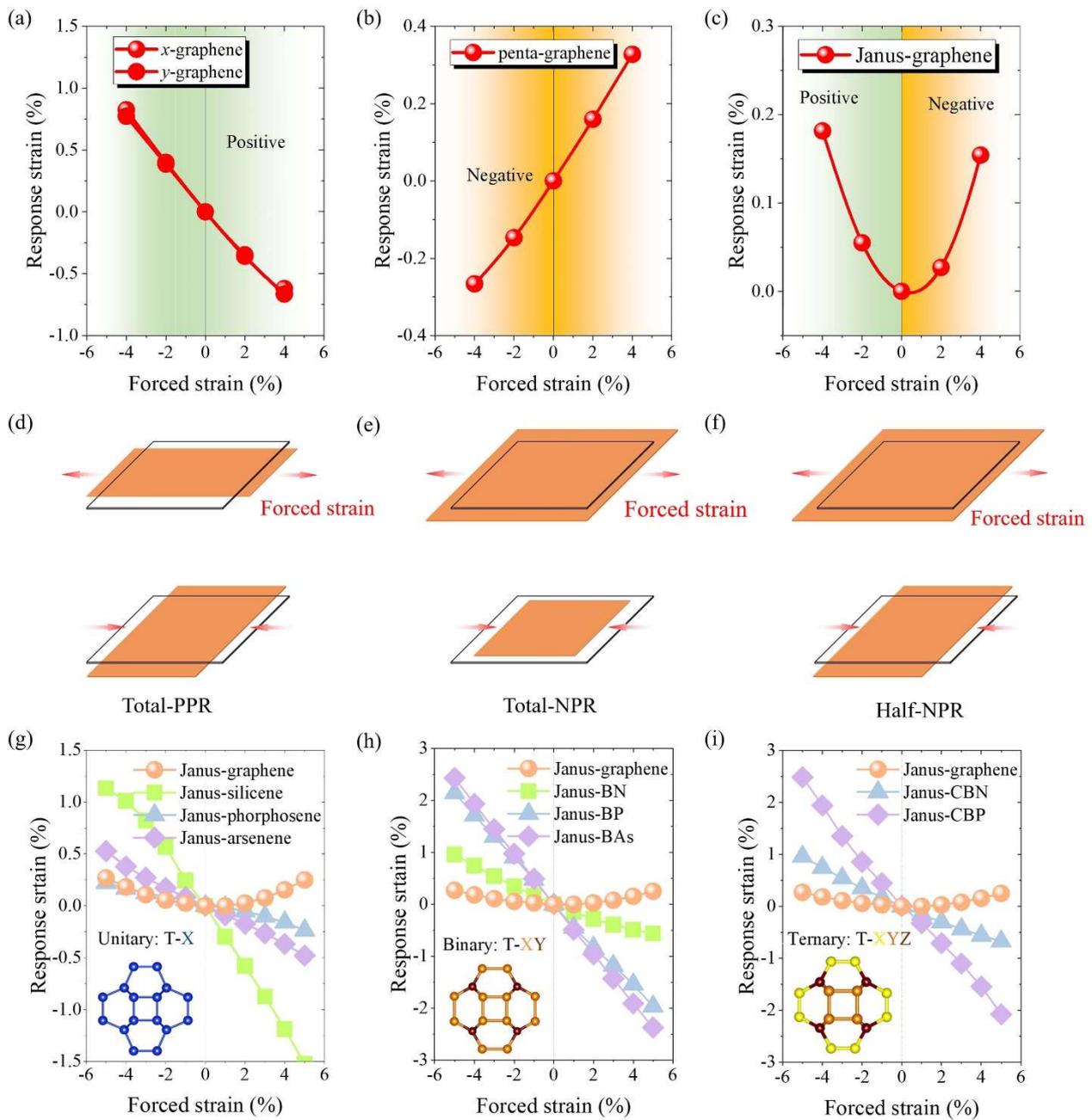

Figure 2. Auxetic behavior of two-dimensional carbon. Strain curves and schematic diagrams representing different Poisson's ratio behaviors for (a) graphene with a (d) total positive Poisson's ratio (Total-PPR) behavior, (b) penta-graphene with a (e) total negative Poisson's ratio (Total-NPR) behavior, and (c) Janus-graphene with a (f) half negative Poisson's ratio (Half-NPR) behavior. Comparison of Poisson's ratio behavior of Janus-graphene with other two-dimensional Janus compounds, including (g) unary, (h) binary, and (i) ternary phases.

## 2.4 Origin of half-auxeticity

To further reveal the origin of half-NPR in Janus-graphene, we demonstrate the strained geometric evolution by tracing atomic trajectories as shown in Fig. 3. The response lattice constant in the orthogonal direction depends on the atoms A, B, C, D, and E and their mirror equivalent atoms, including the lengths $r$, $l$, $m$, $n$ and the angles $\alpha$ and $\beta$. Figs. 3 (a-d) reveals the geometric evolution details of Janus-graphene, Janus-BN, and Janus-CBN. When the strain is applied, the lattice constant of Janus-graphene along the orthogonal can be expressed as $a = n + 2r\cos(\pi - \beta)$, where the response strain depends on the bond length $n$ between mirrored atoms D and E, and the projected length $2r\cos(\pi - \beta)$ determined by the changes of the distance $r$ between atoms A and D and interatomic angle $\beta$ formed by the atoms A, D and E.

To further understand the geometrical evolution of Janus-graphene's half-auxetic behavior, detailed analysis is performed based on 3D and planar 2D schematic diagrams as presented in Fig. 3(e). When the structures are compressed, all Janus materials exhibit expansion behavior consistent with normal materials (*i.e.*, PPR [Fig. 3(e1)]). However, Janus-graphene still expands (*i.e.*, NPR [Fig. 3(e1)]) when stretched, whereas other Janus materials shrink (*i.e.*, PPR [Fig. 3(e2)]). Interestingly, the evolution direction of the geometric parameters of different Poisson's ratio behaviors remains consistent, *i.e.*, $n$, $r$ → "-", and $\beta$ → "+" (due to the same crystal configuration). Note that increasing angle $\beta$ contributes to NPR behavior by inducing an increase in $\cos(\pi - \beta)$, while decreasing $n$, $r$ leads to PPR behavior, revealing a competitive effect. As shown in Fig. 3(d), a stronger β response is captured in Janus-graphene, indicating a dominant role in the competition. Therefore, only Janus-graphene exhibits NPR behavior when stretched.

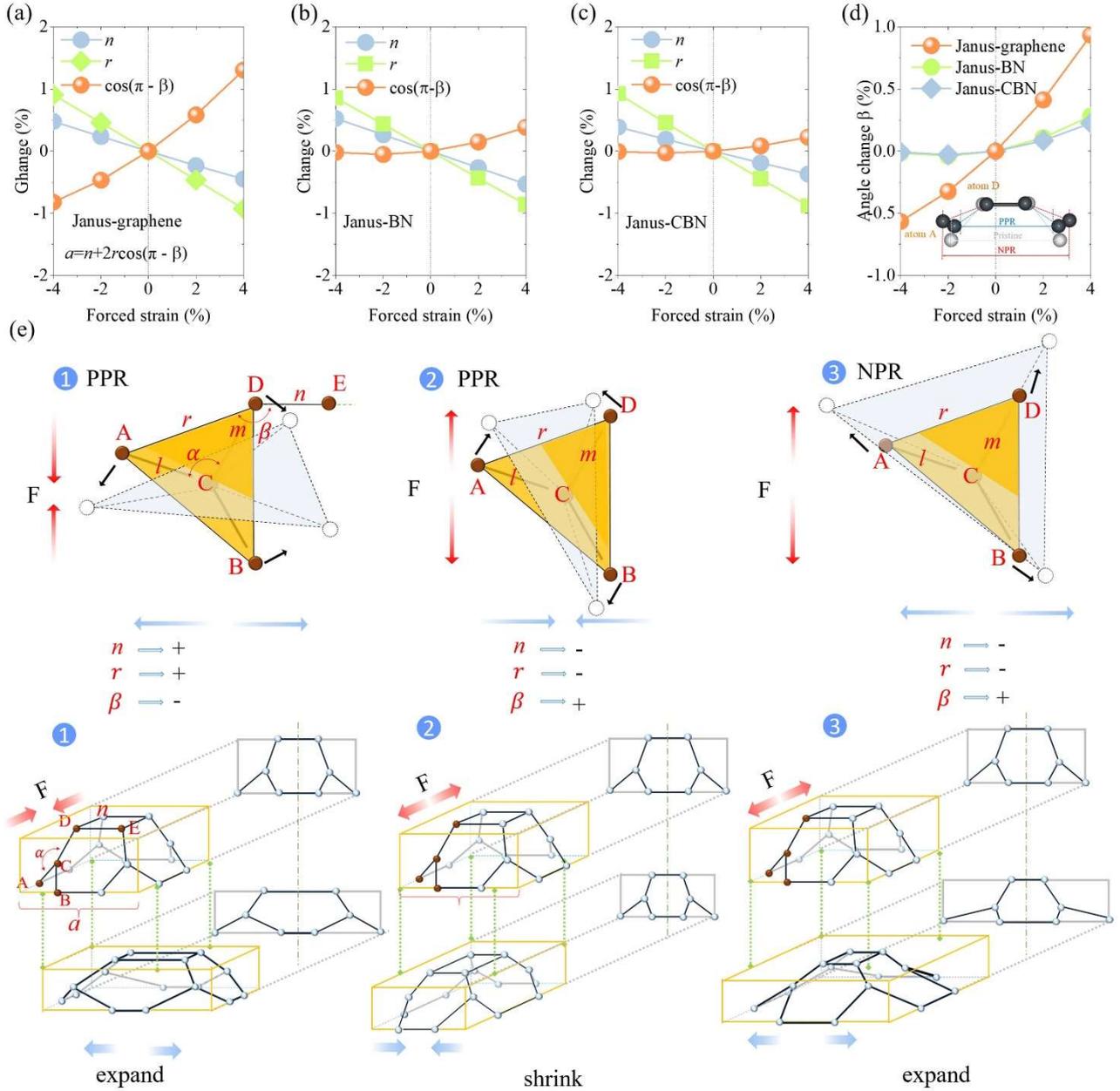

Figure 3. Evolution of Negative Poisson's Ratio. (a-d) The relevant geometric parameters vary with strain. (e) Symmetric atomic configurations and local configuration diagrams in the evolution of negative Poisson's ratio.

## 2.5 Electronic properties and the role of the $p_z$ orbital

Before closing, the electronic properties of Janus-graphene are investigated to further understand the emerging features from the intrinsic Janus structure, as shown in Fig. 4. The band structure of Janus-graphene is plotted in Fig. 4(a) for comparison with graphene in Fig. 4(b). The valence band maximum (VBM) and the conduction band minimum (CBM) of graphene meet at point K to form the well-known Dirac cone, where $p_z$ orbitals form a ring-like uniform large π bond. However, the zero bandgap makes graphene-based field-effect transistors (FETs) unable to be turned on and off by gate control and cannot be applied in the electronics industry. Due to the open $p_z$ orbital, the VBM of Janus-graphene is located at point Γ in the Brillouin zone, while the CBM is located at point M, demonstrating semiconducting properties with an indirect band gap. The *s* orbital electrons mainly contribute to the energy level below -8 eV [Fig.4 (c)], and $p_x$ [Fig. 4(d)] and $p_y$ [Fig. 4(e)] mainly contribute to the energy level range from -4 to -12 eV, revealing the role of $p_z$ orbitals in opening the bandgap.

Further, the precise HSE06 functional identifies a wide bandgap of 2.21 eV for Janus-graphene, higher than the 1.16 eV of the PBE functional. Compared with graphene, the VBM and the CBM mainly contributed by the $p_z$ orbital are not at the same high symmetry point [Fig. 4(f)], which can be further revealed by 3D images of VBM versus CBM as shown in Fig. 4(g). Thus, the deviation of the out-of-plane π bonds in space in the vertical direction leads to the breaking of out-of-plane symmetry.

In the planar phase, the intrinsic Poisson's ratio is generally positive because the electrons of the $p_z$ orbitals are coupled in the large ring-shaped π bonds and cannot be easily activated by the force field to generate a re-entrant mechanism due to the robust π bonds and symmetric planar structure. Typically, planar graphene often needs to apply a strain field of more than 15% to activate the reentrant mechanism to achieve NPR[78,79], which is usually unacceptable in practical operations. As shown in Fig. 4(i), the covalent electrons in the C-C bond in Janus-graphene are unevenly distributed along the out-of-plane direction, *i.e.* strong asymmetry. This leads to intrinsic Janus properties in the structure and also makes the re-entry mechanism easier to activate due to the more isolated $p_z$ electrons. As mentioned in Sec. 2.4, during the structure evolution, the $p_z$ electrons in Janus-grpahene deviate from the planar state and become more sensitive to the external force field. Under a stretching force field, it is forced to couple in the in-plane direction, producing an anomalous response in geometrical angles (β response), exciting potential re-entry mechanisms, and finally leading to NPR. The unique orbital evolution of Janus-graphene brings it non-chemical Janus properties and unconventional NPR effects in terms of geometric and mechanical responses, making it an excellent platform for studying structural and functional linkages.

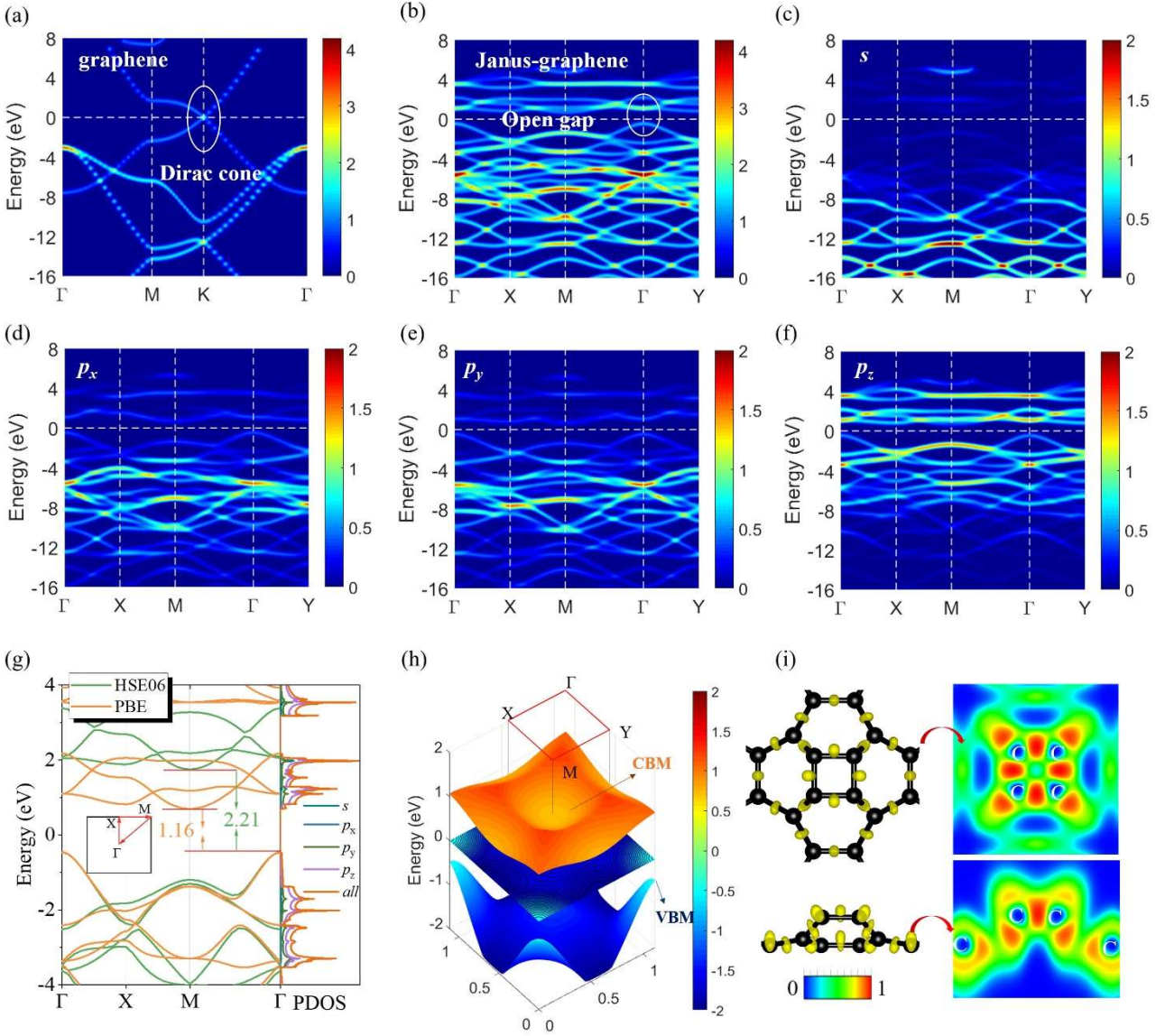

Figure 4. Electronic properties. (a) The band structure of Janus-graphene. (b) The band structure of graphene. The (c) *s*, (d) $p_x$, (e) $p_y$, and (f) $p_z$ orbital electrons occupy the energy bands in Janus-graphene. (g) Electronic band structure of Janus-graphene. (h) Three-dimensional (3D) valence band maximum (VBM) and the conduction band minimum (CBM) in Janus-graphene. The (i) 3D and 2D electron localization function (ELF) of Janus-graphene.

## 3. Conclusion

In summary, we engineered a pure $sp^2$ hybridized 2D carbon allotrope based on carbon gene recombination. This carbon allotrope exhibits double-faced features similar to the Roman god Janus, hence named as Janus-graphene. Different from the traditional Janus configuration, the Janus configuration of Janus-graphene does not depend on chemical effects, which is formed by the carbon atoms growth towards the out-of-plane unilateral side of the 2D plane. The unique Janus geometry maps unconventional half-auxetic property that the Janus-graphene always expands whether it is stretched or compressed. This half-auxetic property shows structure-independence feature, which has never been previously observed in carbon structures, especially in pure $sp^2$-hybridized carbon monolayers. We extend this auxeticity to pure $sp^2$-hybridized configurations and identify its origin in the structure's diverse response to strain. The unique Janus structure and unconventional half-auxetic properties combined with wide-bandgap semiconducting properties endow Janus-graphene with multifunction, providing a strong candidate for novel micro-nanoelectronic devices.

## 4. Computational Methodology

Accurate energy and force calculations are based on density functional theory (DFT) by the Perdew–Burke–Ernzerhof (PBE)[80] functional with the *Vienna ab initio simulation package* (VASP)[81] code. The kinetic energy cutoff of 1000 eV with a 21×21×1 Monkhorst-Pack[82] *q*-mesh was used for structure optimization until the energy and the Hellmann-Feynman force accuracy are $10^{-6}$ eV and $10^{-4}$ eV/Å. The phonon dispersion is calculated based on the finite displacement difference method via the PHONOPY code[83] with the 3×3×1 supercell. *Ab initio molecular dynamics* (AIMD) are calculated based on the canonical ensemble based on VASP code. Poisson's ratio is the negative ratio of the transverse strain $\nu_b$ to the applied axial strain $\nu_a$ of a material when it is stretched or compressed unidirectionally, *i.e.*, $\nu_{ab} = -\nu_b/\nu_a$. In calculating Poisson's ratio, we fix the lattice constant for the applied strain, while fully relaxing the lattice in the orthogonal direction to obtain the transverse strain.

## Acknowledgments

This work is supported by the National Natural Science Foundation of China (Grant No. 52006057), the Natural Science Foundation of Chongqing, China (No. CSTB2022NSCQ-MSX0332), the Fundamental Research Funds for the Central Universities (Grant Nos. 531119200237), and the State Key Laboratory of Advanced Design and Manufacturing for Vehicle Body at Hunan University (Grant No. 52175013). H.W. is supported by the National Natural Science Foundation of China (Grant No. 51906097). Z.Q. is supported by the National Natural Science Foundation of China (Grant No.12274374, 11904324) and the China Postdoctoral Science Foundation (2018M642776). The numerical calculations in this paper have been done on the supercomputing system of the E.T. Cluster and the National Supercomputing Center in Changsha.

## AUTHOR CONTRIBUTIONS

*G.Q.* supervised the project. *L.Y.* performed all the calculations, analysis and writing. All the authors contributed to interpreting the results. The manuscript was written by *L.Y.* with contributions from all the authors.

## COMPETING INTERESTS

The Authors declare no Competing Financial or Non-Financial Interests

## DATA AVAILABILITY

The data that support the findings of this study are available from the corresponding author on reasonable request.